\documentstyle[sprocl]{article}

\def\be{\begin{equation}}
\def\ee{\end{equation}}
\def\bea{\begin{eqnarray}}
\def\eea{\end{eqnarray}}

\def\etal{{\rm et~al. }}
\def\hmpc{\;h^{-1}{\rm Mpc}}
\def\kms{{\rm \;km\;s^{-1}}}
\def\lya{Ly$\alpha$}

\begin{document}

\title{Cosmology from the structure of the \lya~forest}

\author{Rupert A.C. Croft, David H. Weinberg }

\address{Dept. of Astronomy, Ohio State University, Columbus,\\ OH
43202, USA.}
\author{ Neal Katz}
\address{Dept. of Physics and Astronomy, University of Massachusetts, Amherst\\
MA 01003, USA.}
\author{ Lars Hernquist}
\address{Lick Observatory, University of California, Santa Cruz,\\ 
CA 95064, USA.}

\maketitle

\abstracts{A convincing physical picture 
for the \lya\ forest has emerged from simulations and 
related semi-analytic studies of structure formation models.
Observations can be be used in the context of
this picture to study cosmology using the structure of the forest.
With the availability of well motivated predictions, not only
has it become possible to test models directly,  but the physical processes
involved appear to be simple enough that we can attempt to
reconstruct aspects of the  underlying cosmology from observations.
We briefly summarise the method of Croft \etal (1997) [1] for 
recovering the primordial mass power spectrum from \lya\ forest data,
emphasising the physical reasons that the derived P(k) is independent
of unknown ``bias factors''. We present an illustrative application
of the method to four quasar ``spectra'' reconstructed from published
line lists.
}

\section{Introduction}
Observations of structure in the Universe contain cosmological information.
This realisation has been the driving force behind much work done in mapping
the distribution of galaxies, statistical analyses of which have been used 
in attempts to estimate $\Omega$, measure the power spectrum of matter
fluctuations, constrain the nature of dark matter, and so on.
This has been done in the context of
 the theory of structure formation by gravitational
 amplification of small initial perturbations, but before the existence
of any detailed predictive ``theory of galaxy formation''.
This usually only allows for an interpretation
of observations  which includes unknown free parameters, most often
involving the relationship between galaxy and mass fluctuations.
Recent work$^{2,3,4,5,6}$
 has shown that the  gravitational instability
theory first invoked to explain the distribution of galaxies also 
naturally makes
 predictions for  the existence and nature  of QSO absorption
phenomena. The physical processes responsible for most QSO absorption
at high redshift (here we will concentrate on the  \lya~ forest
 caused by neutral hydrogen) appear to be simple enough
that we can understand and simulate them reliably. Therefore,
a ``theory of \lya~ forest formation'' does exist, which can be
taken advantage of to study cosmology in a way which can only be done with 
the galaxy distribution  once there is a ``theory of galaxy formation''.

This contribution to the proceedings concerns one optimistic 
way of viewing the situation,  namely that if the picture we have for the
formation of the \lya\ forest is broadly
correct, then we can ask what this entails for our cosmology. 
The primordial power spectrum of matter fluctuations, P(k) is of particular 
interest. We will briefly summarise a method (details are in [1])
for estimating this quantity and give some illustrative results.

\section{Reconstruction of P(k) from the \lya~ forest}
To reconstruct P(k), a statistic based on the mass distribution, we
first need to understand how the \lya~ forest of absorption is related
to the mass distribution in our chosen theoretical picture,
the gravitational instability scenario for
structure formation. Hydrodynamic simulations and 
semi-analytic work have revealed that the high-$z$ \lya\ forest 
arises in  the large fraction of space where
the density of matter is within a factor of 10 of the cosmic mean.
In these regions, pressure effects are relatively unimportant, so that the
gas tends to trace structure in the dark matter.
The local density of  matter also governs the temperature of the gas, which
follows a power law temperature-density relation$^{7}$, $T=T_{0}\rho^{0.6}$,
where $\rho$ is the density in units
of the cosmic mean. This relation arises because of the interplay between
photoionization heating by the UV background and adiabiatic cooling by
the expansion of the Universe.
If we ignore other effects for the moment, such as thermal broadening,
peculiar motions and collisional ionisation, we can infer that 
the optical depth for absorption ($\tau$) and the local density 
obey the following approximate relation:
\begin{eqnarray}
&\tau \propto n_{HI}~= A~{\rho}^{\beta}, &
\label{eqn:tau} \\
&A = 0.946 \left(\frac{1+z}{4}\right)^6 
\left(\frac{\Omega_b h^2}{0.0125}\right)^2
\left(\frac{T_0}{10^4\;{\rm K}}\right)^{-0.7}
\left(\frac{\Gamma}{10^{-12}\;{\rm s}^{-1}}\right)^{-1}
\left(\frac{H(z)}{100\;\kms\;{\rm Mpc}^{-1}}\right)^{-1}, &
\nonumber
\end{eqnarray}
where $n_{HI}$ is the neutral hydrogen density.  
 $\beta$ is  $\sim 1.6$ 
and the parameter $A$ is a function of the baryon
density ($\Omega_{b}$), the photoionization rate ($\Gamma$),$T_{0}$,
$z$ and H$_{0}$. 
 The flux observed at a point in the QSO spectrum,
$F$,  is given by $e^{-\tau}$.
This relation between flux and mass is analogous to the local biasing 
relations which have been postulated to exist between galaxies and mass.
The difference here is that it is a prediction directly
derived from theory, and that the parameter $A$ can be inferred from
observations, as will be explained below.

As the relation between flux and mass is local, we would
expect the shape of the power spectrum of the flux to be the same above
some scale as the shape of P(k) for the mass. In tests on simulations, we find
that this is indeed the case for $\lambda > 1.5 \hmpc$. In our method
we therefore recover the shape of P(k) for the mass from the power 
spectrum of the flux. There are 
two additional twists that we incorporate at this
stage. The first involves the fact that by taking an FFT of a QSO spectrum 
we measure the power in 1 dimension. As we are interested in P(k) in 3-d
, we must convert the quantity to 3-d, which involves as simple
differentiation (see [1], equation 5). The second is that as we are
interested in the initial, linear P(k) of the mass,  we apply a
monotonic transformation to the flux, to give it a Gaussian pdf, before
measuring P(k). This step is motivated by the fact that under gravitational
instability, the rank order of (smoothed) 
densities is seen to be approximately
conserved$^{8}$, so that one way of recovering the
initial density field is to monotonically map the final densities back to the
initial pdf$^{8}$, here assumed to be Gaussian.
As this is a local transformation, it does not alter the shape of
P(k) for  $\lambda > 1.5 \hmpc$, but it 
does have the effect of slightly reducing the noise level. 

 We now have the shape of the mass P(k), but not its amplitude. To find
this normalisation, we use the fact that a higher amplitude of mass 
fluctuations will result in larger fluctuations in the flux.
We can therefore run simulations set up using the 
mass P(k) shape we have derived but with different mass amplitudes,
and pick the mass amplitude which yields the observed level of flux
fluctuations in the simulated spectra
we extract. The measure of flux fluctuations 
we use is the power spectrum of the flux.
In carrying out this procedure, we must bear in mind that a number of
other factors may also influence the amplitude of flux fluctuations.
The first of these is obviously the parameter $A$ in equation (1), as a larger
value of $A$ will give larger flux fluctuations for a given 
set of $\rho$ (mass) fluctuations. Luckily, as mentioned above, the
value of $A$ can actually be fixed using a measurement
made independently from the observations. This measurement is the mean
flux level $\left<F\right>$ in the observations, which, for a given set
of mass fluctuations, will specify $A$ uniquely. Being able to make this
measurement is analogous to being able to measure the mass associated with
each galaxy, which would then specify galaxy biasing exactly.
Once we have fixed $A$, we have to pick the other parameters we need 
to run the normalizing simulations, such as the value of $\Omega$, the mean
temperature of the gas, and so on. Varying these parameters has little
effect on the results, basically because the approximation of 
equation (1) is a good one on the scales where we are interested in
measuring P(k). 

For speed, our normalising simulations are run not with a full hydro
code, but with a PM N-body code, assuming that gas and dark matter
follow each other closely. We have compared this approach to the 
full hydro case and find that it works very well for our purpose.
We now demonstrate the application of our P(k) reconstruction method
to observations, both simulated and real.

\section{Illustrative results}

\begin{figure}[t]
\vspace{4.4cm}
\includegraphics{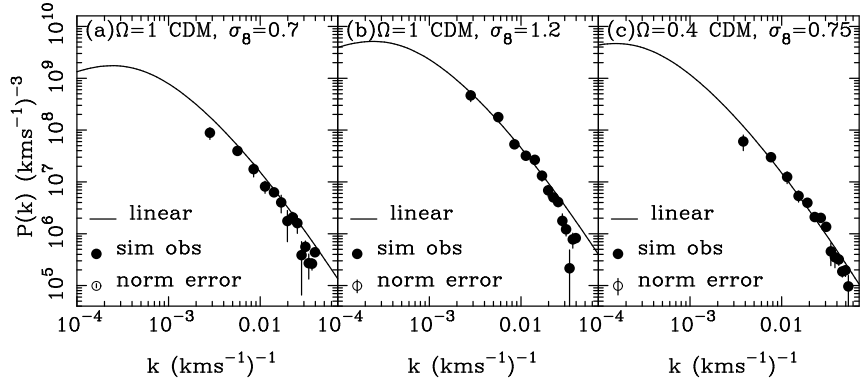}
\caption[]{
The mass power spectrum, P(k),
recovered from  simulated observations
of the \lya\ forest at $z=3$ (see text). The simulations used were of three
different CDM models, with parameters listed at the top of each panel.
$H_{0}$ was $50 \kms\rm{Mpc}^{-1}$ for (a) and (b) and 
$65\kms\rm{Mpc}^{-1}$ for (c). An  estimate 
of the error in the overall normalisation
of the recovered (solid) points is shown in the bottom left.
The true linear P(k) for each model is shown as a solid line.}
\end{figure}

We test the method by using full hydrodynamic simulations run
with the TreeSPH code$^{4}$, which includes gas pressure,
collisional ionisation,  shock heating, star formation and other
 effects not explicitly included in the approximate description listed above.
We use these simulations (of three different CDM models) to create 
QSO spectra, which we rebin into coarse (40 $\kms$) pixels
and then add Possion photon noise (S/N=10) to, in order to simulate
medium quality (not state of the art) observations.
 These simulated spectra (for $z=3$) are then 
analysed with the P(k) reconstruction method,
using enough lines of sight through the simulation box to be equivalent 
 in length  to 5 QSO spectra. The results are shown in
Figure 1, together with a curve showing the correct linear theory P(k)
for each model. These results are displayed in the relevant observational
units, $\kms$, but in each case, the largest scale point corresponds to
the largest mode in the simulation box, which has
 wavelength $\lambda=11.11 \hmpc$.
Figure 1 shows that we can recover P(k) reasonably well
on scales from $\sim 1-10 \hmpc$.

\begin{figure}[t]
\vspace{5.4cm}
\includegraphics{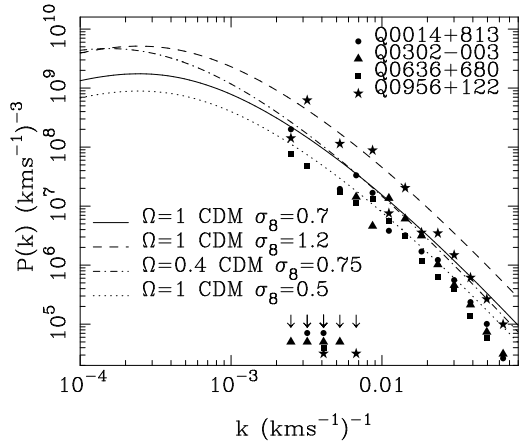}
\caption[]{
Symbols show the mass power spectra recovered from the \lya\ forest
 in four QSO spectra
reconstructed on the basis of published line lists (taken from
[9]). 
The linear theory power spectra for different CDM models
at the same mean redshift $z=2.9$ are also shown.}
\end{figure}

In advance of an application to a large sample of observational
data, we have obtained
some illustrative results by applying the method to some published data 
from [9]
for 4 quasars, also of absorption at  $z\sim 3$. The published data
is in the form of lists of lines with Voigt profiles into which the spectra
have been decomposed. We have reconstructed the spectra on the basis of these
lists, as in our theoretical picture we need a continuous flux distribution
which we relate to the continuous mass distribution. There may be unknown
systematics involved in this reconstruction from line lists, so it should be
borne in mind that the results which we show (in Figure 2)  are for
illustrative purposes only. We can see that the reconstructed P(k),
whilst being consistent with low amplitude CDM models exhibits large 
scatter from quasar to quasar.
In some cases, the differentiation required to convert 1-d power to 3-d power 
yields a negative P(k) because of noise; in such cases we plot a point at the
base of the figure.
 There exist large samples of data, most 
at lower resolution, which we hope to use soon in a definitive reconstruction
of P(k).

\section{Discussion}
So far we have assumed that the photoionising UV background is 
completely uniform, when in fact it is probably 
 produced by discrete sources such as QSOs and
Population III stars . This would cause fluctuations in 
the local value of the photoionisation rate which 
enters into equation (1) and so cause fluctuations in the flux
on top of those associated with the value of $\rho$ at that point.
Other effects could also cause inhomogenous heating
of the gas  which could also affect the accuracy of our 
approximations. It would be surprising, however, if any of these effects
were strong enough to change our P(k) measurement substantially. This
is because on the scales we are  estimating P(k), the fluctuations caused by
the mass are of order unity, and any competing physical effect would have
to be of similar strength and occur over most of the QSO spectrum to make
a difference. A rough estimate (made in the optically thin limit)
 of the amplitude of fluctuations 
expected to be caused by QSO discreteness was made in [10].
It was  calculated that on the largest scales we measure P(k) here
, the fluctuations in the background would make a contribution to P(k)
 of about $2 \%$ of that expected to come from the mass. A higher density
of sources should in principle cause smaller fluctuations. We can
be reasonably confident that more detailed simulations
will not change these preliminary conclusions.

Apart from the reconstruction of P(k), the theoretical picture for the 
origin of the \lya\ forest  used in this paper has already been used to
constrain the baryon density$^{11,12}$ and directly
test cosmological models$^{11}$.
 Particularly with an
extension to three dimensions by cross-correlating the spectra of nearby
quasars, it seems as though there should be  many other 
ways that the \lya\ forest can be used to robustly constrain 
our cosmology.

\end{document}